\documentstyle[amsfonts,preprint,aps]{revtex}

\begin{document}
\title{Electromagnetic diffraction by a circular cylinder with longitudinal slots}
\author{B. Guizal and D.\ Felbacq}
\address{LASMEA UMR 6602\\
Complexe des C\'{e}zeaux\\
63177 Aubi\`{e}re Cedex\\
France}
\maketitle

\begin{abstract}
A method is presented to investigate diffraction of an electromagnetic plane
wave by an infinitely thin infinitely conducting circular cylinder with
longitudinal slots. It is based on the use of the combined boundary
conditions method that consists on expressing the continuity of the
tangential components of both the electric and the magnetic fields in a
single equation. This method proves to be very efficient for this kind of
problems and leads to fast numerical codes.
\end{abstract}

\section{Introduction}

\noindent \tightenlines The problem of the penetration of electromagnetic
waves in a conducting circular cavity through a narrow axial aperture has
been treated by several authors.\ Several methods have been used to achieve
the determination of the field inside the cavity.\ Beren \cite{beren} used
the Aperture Field Integral Equation , the Electric Field Integral Equation
and H-field Integral Equation to determine the field around an axially
slotted cylinder, while Johnson and Ziolkowski \cite{johnson}\ gave a
generalized dual series solution for this problem. Mautz and Harrington
treated the field penetration inside a conducting circular cylinder through
a narrow slot in both TE \cite{mautz1}\ and TM \cite{mautz2} polarizations.
More recently Shumpert and Butler \cite{shump1}, \cite{shump2} proposed
three methods to study the penetration in conducting cylinders. In this
article, we propose a method to calculate the field inside and around a
slotted circular cavity with longitudinal slots. It is based on the combined
boundary conditions method introduced first by Montiel and Nevi\`{e}re \cite
{montiel,montiel2}. Section II is dedicated to the description of the
theory. In section III we give some details about the numerical scheme and
then compare our results with previous work.

\section{Theory}

\noindent The structure under study is depicted in Fig 1. The space is
divided into two regions, region 1 (exterior region : $r>R$) \ and region 2
(interior : $r<R$) that are assumed to be dielectric and homogeneous with
relative dielectric permittivities $\varepsilon _{1}$ and $\varepsilon _{2}$%
, respectively. On the interface between these two media are deposited a
finite number of infinitely conducting, infinitely thin circular strips that
are invariant along the $z$ direction. The device is illuminated by a TMz
(electric field parallel to the $z$ axis) or TEz (magnetic field parallel to
the $z$ axis) monochromatic electromagnetic wave under incidence $\theta
_{0} $\ with vacuum wavelength $\lambda $.\newline
Throughout this paper we assume an $\exp \left( -i\omega t\right) $ time
dependence. The $z$ component of the electric or the magnetic field will be
denoted by $u\left( \theta ,r\right) .$ We denote by $\Omega _{1}$ the union
of the strips and by $\Omega _{2}$ its complementary in $\left[ 0,2\pi %
\right] .$ \newline
In the exterior region we express the total field as :

\begin{equation}
u_{1}\left( \theta ,r\right) =%
\mathrel{\mathop{\stackrel{}{\sum }}\limits_{n\in {\Bbb Z}}}%
a_{n}J_{n}(k_{1}r)\exp (in\theta )+%
\mathrel{\mathop{\sum }\limits_{n\in {\Bbb Z}}}%
b_{n}H_{n}^{\left( 1\right) }(k_{1}r)\exp (in\theta )  \label{eq1}
\end{equation}
Likewise in the interior region the total field may be expressed as :

\begin{equation}
u_{2}\left( \theta ,r\right) =%
\mathrel{\mathop{\sum }\limits_{n\in {\Bbb Z}}}%
c_{n}J_{n}(k_{2}r)\exp (in\theta )  \label{eq2}
\end{equation}
where $a_{n}$, $b_{n}$\ and $c_{n}$ are the amplitudes of the incident, the
diffracted and the transmitted waves respectively. We denote $J_{n}$ and $%
H_{n}^{\left( 1\right) }$ the Bessel and the Hankel functions of the first
kind. $k_{p}=k_{0}\sqrt{\varepsilon _{p}}=%
{\displaystyle{2\pi  \over \lambda }}%
\sqrt{\varepsilon _{p}},$ with $\ p=1,2$ and ${\Bbb Z}$ denotes the usual
set of relative integers.

\noindent Amplitudes $a_{n}$ being known, the problem is to determine
amplitudes $b_{n}$ and $c_{n}$ from which the total field can be calculated
everywhere. For that purpose one must write the boundary conditions at\ the
interface between both dielectric media. This is done in the next
subsections by distinguishing the TMz and the TEz cases of polarization.

\subsection{TMz polarization}

\noindent The boundary conditions applied to the tangential components of
the electromagnetic field at the interface defined by $r=R$\ lead to :

\begin{mathletters}
\begin{eqnarray}
&&\left. u_{1}\left( \theta ,R\right) =u_{2}\left( \theta ,R\right) ,\text{ }%
\forall \theta \in \left[ 0,2\pi \right] \right.  \label{eq3} \\
&&\left( 
{\displaystyle{du_{1} \over dr}}%
\right) _{\left( \theta ,R\right) }=\left( 
{\displaystyle{du_{2} \over dr}}%
\right) _{\left( \theta ,R\right) },\text{\ }\forall \theta \in \Omega _{2}%
\text{ }  \label{eq33}
\end{eqnarray}
With the supplementary condition that the electric field must vanish on the
strips:

\end{mathletters}
\begin{equation}
u_{1}\left( \theta ,R\right) =u_{2}\left( \theta ,R\right) =0,\forall \theta
\in \Omega _{1}  \label{eq4}
\end{equation}
Following Montiel and Nevi\`{e}re\cite{montiel,montiel2}, equations (\ref
{eq33}) and (\ref{eq4}) can be combined in a single equation that holds for
every $\theta $ in $\left[ 0,2\pi \right] $ :

\begin{equation}
\left( 1-\chi \left( \theta \right) \right) u_{2}\left( \theta ,R\right)
+g\chi \left( \theta \right) \left[ \left( 
{\displaystyle{du_{2} \over dr}}%
\right) _{\left( \theta ,R\right) }-\left( 
{\displaystyle{du_{1} \over dr}}%
\right) _{\left( \theta ,R\right) }\right] =0,\text{ \ }\forall \theta \in %
\left[ 0,2\pi \right]  \label{eq5}
\end{equation}
where $\chi \left( \theta \right) $ is the characteristic function of set $%
\Omega _{2}$: 
\[
\chi \left( \theta \right) =\left\{ 
\begin{array}{ll}
1 & \text{if \ }x\in \Omega _{2} \\ 
0 & \text{elsewhere}
\end{array}
\right. 
\]
and $g$ is some numerical parameter introduced for dimensional and numerical
purposes. Remark that the set of Eqs. (\ref{eq3}), (\ref{eq33}) and (\ref
{eq4}) is equivalent to the set of Eqs.\ (\ref{eq3}) and (\ref{eq5}). Since $%
\chi \left( \theta \right) $ is $2\pi $-periodic it can be expanded in
Fourier series$:$

\begin{equation}
\chi \left( \theta \right) =%
\mathrel{\mathop{\sum }\limits_{p\in {\Bbb Z}}}%
\chi _{p}\exp (ip\theta )  \label{eq6}
\end{equation}
Reporting equations (\ref{eq1}) and (\ref{eq2}) into equation (\ref{eq3})
and projecting on the $\left( \exp (in\theta )\right) _{n\in {\Bbb Z}}$\
basis gives :

\begin{equation}
a_{n}J_{n}(k_{1}R)+b_{n}H_{n}^{\left( 1\right)
}(k_{1}R)=c_{n}J_{n}(k_{2}R),\qquad \forall n\in {\Bbb Z}  \label{eq7}
\end{equation}
then reporting equations (\ref{eq1}),(\ref{eq2})\ and (\ref{eq6}) into
equation (\ref{eq5}) and projecting on the $\left( \exp (in\theta )\right)
_{n\in {\Bbb Z}}$\ basis leads to :

\begin{equation}
\left. 
\begin{array}{l}
c_{n}J_{n}(k_{2}R)-%
\mathrel{\mathop{\sum }\limits_{p\in {\Bbb Z}}}%
\chi _{n-p}c_{p}J_{p}(k_{2}R)+ \\ 
\qquad \qquad \qquad g%
\mathrel{\mathop{\sum }\limits_{p\in {\Bbb Z}}}%
\chi _{n-p}\left[ k_{2}\left( c_{p}J_{p}^{\prime }(k_{2}R)\right)
-k_{1}\left( a_{p}J_{p}^{\prime }(k_{1}R)+b_{p}H_{p}^{\left( 1\right) \prime
}(k_{1}R)\right) \right] =0,\qquad \forall n\in {\Bbb Z}
\end{array}
\right.  \label{eq8}
\end{equation}
\newline
where the primes denote derivation with respect to $r.$ From Eq(\ref{eq7})
one can extract $c_{n}$ :

\begin{equation}
c_{n}=\frac{J_{n}(k_{1}R)}{J_{n}(k_{2}R)}a_{n}+\frac{H_{n}^{\left( 1\right)
}(k_{1}R)}{J_{n}(k_{2}R)}b_{n},\qquad \forall n\in {\Bbb Z}  \label{eq9}
\end{equation}
and report its expression into Eq(\ref{eq8}) to obtain the following linear
system linking the amplitudes $a_{n}$ and $b_{n}:$

\begin{equation}
\left. 
\begin{array}{l}
a_{n}J_{n}(k_{1}R)+%
\mathrel{\mathop{\sum }\limits_{p\in {\Bbb Z}}}%
\chi _{n-p}a_{p}\left[ -J_{p}(k_{1}R)+gk_{2}%
{\displaystyle{J_{p}(k_{1}R) \over J_{p}(k_{2}R)}}%
J_{p}^{\prime }(k_{2}R)-gk_{1}J_{p}^{\prime }(k_{1}R)\right] = \\ 
-b_{n}H_{n}^{\left( 1\right) }(k_{1}R)+%
\mathrel{\mathop{\sum }\limits_{p\in {\Bbb Z}}}%
\chi _{n-p}b_{p}\left[ H_{p}^{\left( 1\right) }(k_{1}R)-gk_{2}%
{\displaystyle{H_{p}^{\left( 1\right) }(k_{1}R) \over J_{p}(k_{2}R)}}%
J_{p}^{\prime }(k_{2}R)+gk_{1}H_{p}^{\left( 1\right) \prime }(k_{1}R)\right]
\end{array}
\right.  \label{eq10}
\end{equation}
The solution of the linear system (\ref{eq10}) gives the unknown amplitudes $%
b_{n}$ and then Eq.(\ref{eq9}) gives the amplitudes $c_{n}.$ Thus the field
can be computed everywhere in space using Eqs.(\ref{eq1}) and (\ref{eq2}).

\subsection{TEz polarization}

\noindent For this case of polarization, the continuity of the tangential
components of the electromagnetic field at the interface defined by $r=R$\
leads to :

\begin{mathletters}
\begin{eqnarray}
&&%
{\displaystyle{1 \over \varepsilon _{1}}}%
\left( 
{\displaystyle{du_{1} \over dr}}%
\right) _{\left( \theta ,R\right) }=%
{\displaystyle{1 \over \varepsilon _{2}}}%
\left( 
{\displaystyle{du_{2} \over dr}}%
\right) _{\left( \theta ,R\right) },\text{\ }\forall \theta \in \left[
0,2\pi \right]  \label{eq11} \\
&&u_{1}\left( \theta ,R\right) =u_{2}\left( \theta ,R\right) ,\text{ \ }%
\forall \theta \in \Omega _{2}  \label{eq1111}
\end{eqnarray}
With the supplementary condition that the electric field must vanish on the
strips:

\end{mathletters}
\begin{equation}
{\displaystyle{1 \over \varepsilon _{1}}}%
\left( 
{\displaystyle{du_{1} \over dr}}%
\right) _{\left( \theta ,R\right) }=%
{\displaystyle{1 \over \varepsilon _{2}}}%
\left( 
{\displaystyle{du_{2} \over dr}}%
\right) _{\left( \theta ,R\right) }=0,\forall \theta \in \Omega _{1}
\label{eq12}
\end{equation}
Here again we can replace equations (\ref{eq1111}) and (\ref{eq12}) by :

\begin{equation}
\left( 1-\chi \left( \theta \right) \right) 
{\displaystyle{1 \over \varepsilon _{2}}}%
\left( 
{\displaystyle{du_{2} \over dr}}%
\right) _{\left( \theta ,R\right) }+g\chi \left( \theta \right) \left[
u_{2}\left( \theta ,R\right) -u_{1}\left( \theta ,R\right) \right] =0,\text{
\ }\forall \theta \in \left[ 0,2\pi \right]  \label{eq13}
\end{equation}
Reporting equations (\ref{eq1}) and (\ref{eq2}) into Eq.(\ref{eq11}) and
projecting on the $\left( \exp (in\theta )\right) _{n\in {\Bbb Z}}$\ basis
gives :

\begin{equation}
a_{n}J_{n}^{\prime }(k_{1}R)+b_{n}H_{n}^{\left( 1\right) \prime }(k_{1}R)=%
\frac{k_{2}}{k_{1}}%
{\displaystyle{\varepsilon _{1} \over \varepsilon _{2}}}%
c_{n}J_{n}^{\prime }(k_{2}R),\qquad \forall n\in {\Bbb Z}
\end{equation}

\noindent Remark that the set of Eqs (\ref{eq11}), (\ref{eq1111}) and (\ref
{eq12}) are equivalent to the set of Eqs (\ref{eq1111}) and (\ref{eq13}).\
Reporting equations (\ref{eq1}),(\ref{eq2})\ and (\ref{eq6}) into Eq.(\ref
{eq13}) and projecting on the $\left( \exp (in\theta )\right) _{n\in {\Bbb Z}%
}$\ basis leads to :

\begin{equation}
\left. 
\begin{array}{l}
{\displaystyle{k_{2} \over \varepsilon _{2}}}%
c_{n}J_{n}^{\prime }(k_{2}R)-%
{\displaystyle{k_{2} \over \varepsilon _{2}}}%
\mathrel{\mathop{\sum }\limits_{p\in {\Bbb Z}}}%
\chi _{n-p}c_{p}J_{p}^{\prime }(k_{2}R)+ \\ 
\qquad \qquad \qquad g%
\mathrel{\mathop{\sum }\limits_{p\in {\Bbb Z}}}%
\chi _{n-p}\left[ c_{p}J_{p}(k_{2}R)-\left(
a_{p}J_{p}(k_{1}R)+b_{p}H_{p}^{\left( 1\right) }(k_{1}R)\right) \right]
=0,\qquad \forall n\in {\Bbb Z}
\end{array}
\right.  \label{eq15}
\end{equation}
From Eq.(14) one can extract $c_{n}$ :

\begin{equation}
c_{n}=%
{\displaystyle{k_{1} \over k_{2}}}%
{\displaystyle{\varepsilon _{2} \over \varepsilon _{1}}}%
\left( \frac{J_{n}^{\prime }(k_{1}R)}{J_{n}^{\prime }(k_{2}R)}a_{n}+\frac{%
H_{n}^{\left( 1\right) \prime }(k_{1}R)}{J_{n}^{\prime }(k_{2}R)}%
b_{n}\right) ,\qquad \forall n\in {\Bbb Z}  \label{eq16}
\end{equation}
and report its expression into Eq.(\ref{eq15}) to obtain the following
linear system linking the amplitudes $a_{n}$ and $b_{n}:$

\begin{equation}
\left. 
\begin{array}{l}
a_{n}%
{\displaystyle{k_{1} \over \varepsilon _{1}}}%
J_{n}^{\prime }(k_{1}R)+%
\mathrel{\mathop{\sum }\limits_{p\in {\Bbb Z}}}%
\chi _{n-p}a_{p}\left[ -%
{\displaystyle{k_{1} \over \varepsilon _{1}}}%
J_{p}^{\prime }(k_{1}R)+g%
{\displaystyle{k_{1} \over k_{2}}}%
{\displaystyle{\varepsilon _{2} \over \varepsilon _{1}}}%
{\displaystyle{J_{p}(k_{2}R) \over J_{p}^{\prime }(k_{2}R)}}%
J_{p}^{\prime }(k_{1}R)-gJ_{p}(k_{1}R)\right] = \\ 
-b_{n}%
{\displaystyle{k_{1} \over \varepsilon _{1}}}%
H_{n}^{\left( 1\right) \prime }(k_{1}R)+%
\mathrel{\mathop{\sum }\limits_{p\in {\Bbb Z}}}%
\chi _{n-p}b_{p}\left[ 
{\displaystyle{k_{1} \over \varepsilon _{1}}}%
H_{p}^{\left( 1\right) \prime }(k_{1}R)-g%
{\displaystyle{k_{1} \over k_{2}}}%
{\displaystyle{\varepsilon _{2} \over \varepsilon _{1}}}%
{\displaystyle{J_{p}(k_{2}R) \over J_{p}^{\prime }(k_{2}R)}}%
H_{p}^{\left( 1\right) \prime }(k_{1}R)+gH_{p}^{\left( 1\right) }(k_{1}R)%
\right]
\end{array}
\right.  \label{eq17}
\end{equation}
\newline
The Solution of the linear system (\ref{eq17}) gives the unknown amplitudes $%
b_{n}$ and then Eq.(\ref{eq16}) gives the amplitudes $c_{n}.$ Thus the field
can be computed everywhere in space using Eqs.(\ref{eq1}) and (\ref{eq2}).

\section{Numerical results}

\noindent The infinite linear systems (\ref{eq10}) and (\ref{eq17}) are
truncated to a finite size by retaining only $\left( 2N+1\right) $
coefficients and solved to obtain a representation of the field at
truncation order $N$. The convergence of the results has been checked by
increasing integer $N$ and using the usual criteria of energy balance (
optical theorem ) and reciprocity. We have also verified that the boundary
conditions are fulfilled, for instance the nullity of the tangential
electric field on the strips. In all the calculations carried in this paper
we set $g=-10^{-3}$. However, as mentioned in \cite{guizal}, numerical
experiments show that only the sign of $g$ is of importance: the numerical
scheme is more stable with a negative value of $g$. All the computations
reported have been obtained on a Personal Computer ($200$ MHz processor with 
$32$ Mo of RAM), only a\ few seconds are necessary to perform each result
shown here.

\noindent In the following we provide some numerical examples and compare
our results with those obtained in previous works \cite{mautz1}, \cite
{mautz2}, \cite{shump1} and \cite{shump2}.

\noindent In our first example, we consider a circular cavity with a single
longitudinal narrow slot (see Fig. 2) with $\phi =5%
{{}^\circ}%
$ and we compute the interior field on $x$ axis.\ Figures 3 (a) and (b) show
the magnitude of the normalized electric field in both the TM$_{z}$ and the
TE$_{z}$ cases of polarization. It can be seen that our results are in
excellent agreement with those published recently by Shumpert and Butler 
\cite{shump1}, \cite{shump2}, see for instance figure 6 in this last
reference. In the second example, we consider a circular cavity with an
aperture such that $\phi =5%
{{}^\circ}%
.$ In figures 4 (a) and (b) are plotted the normalized electric field
amplitude at the center of the cylinder for various values of the parameter $%
k_{1}R$\ for both the TM$_{z}$ and the TE$_{z}$ cases of polarization. These
curves agree with those obtained by Mautz and Harrington (see references 
\cite{mautz} and \cite{mautzz} ). It is worth noticing that the resonances
in these plots correspond to the modes of the cavity.

\noindent Finally we give the map of the electric field around and inside
the slotted cylinder when excited by a plane wave such that $k_{1}R$
corresponds to a mode of the closed cylinder. We can see in figures 5 (a)
and (b) that the modes TM$_{01}(k_{1}R=2.404)$ and TM$_{11}(k_{1}R=3.832)$
are excited inside the structure.

\section{Conclusion}

\noindent We have developed a very efficient and fast method adapted to
study diffraction of an electromagnetic wave by a finite number of
infinitely thin, infinitely conducting strips deposited on a dielectric
cylinder. It is based on the combined boundary conditions method. The method
is very low CPU-time consuming. The numerical examples that have been given
to illustrate the method are not restrictive. One can use as an incident
radiation a beam of any shape. It suffices to calculate its corresponding
incident amplitudes $a_{n}$. It is also possible to study the radiation
pattern of a source located at the center of the cylinder by making slight
changes in the equations.

{\LARGE Figure captions}\newline

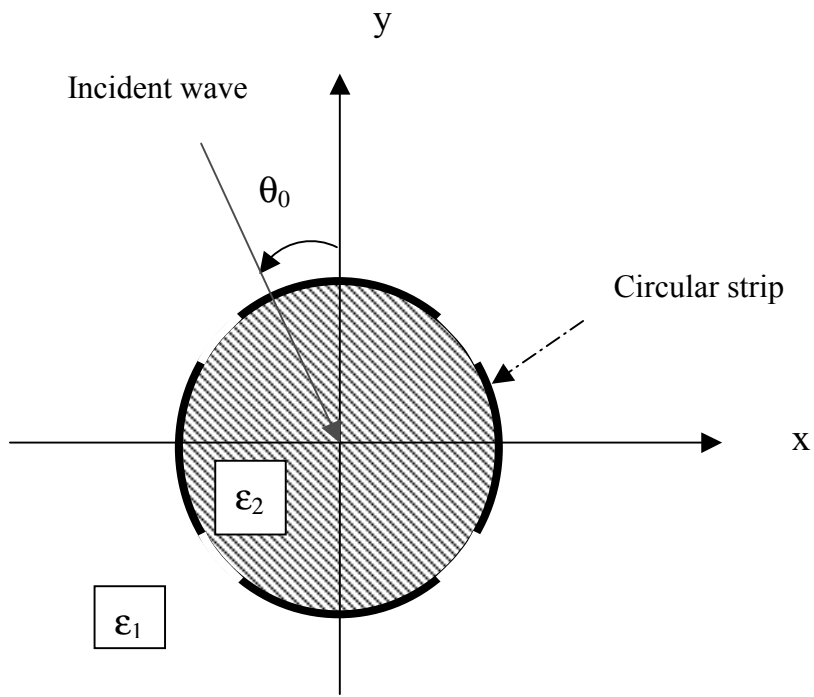
{\bf Figure 1} : Geometry of the problem : a TEz or a TMz polarized plane
wave illuminates

\qquad \qquad\ \ \ the slotted cylinder

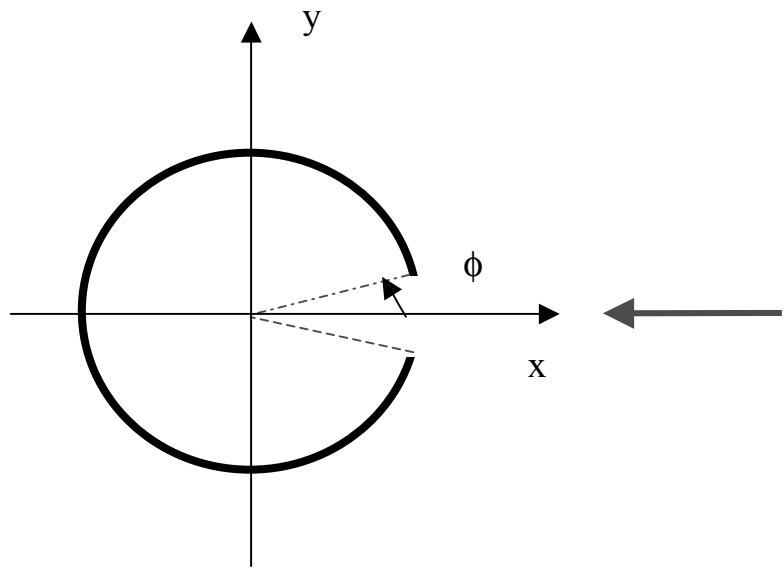
{\bf Figure 2 : }Electromagnetic{\bf \ }penetration into a circular cavity
through a narrow slot.

{\bf Figure 3} : (a) Magnitude of normalized electric field on the x axis of
\ a slotted circular

\ \ \ \ \ \ \ \ \ \ \ \ \ \ \ \ \ \ \ \ cylinder excited by TM$_{z}$ plane
wave $\left\{ k_{1}R=0.7,\theta _{0}=180%
{{}^\circ}%
,\phi =5%
{{}^\circ}%
\right\} $

\ \ \ \ \ \ \ \ \ \ \ \ \ \ \ \ (b) Magnitude of normalized electric field
on the x axis of slotted circular

\ \ \ \ \ \ \ \ \ \ \ \ \ \ \ \ \ \ \ \ cylinder excited by TE$_{z}$ plane
wave $\left\{ k_{1}R=0.7,\theta _{0}=0%
{{}^\circ}%
,\phi =5%
{{}^\circ}%
\right\} $

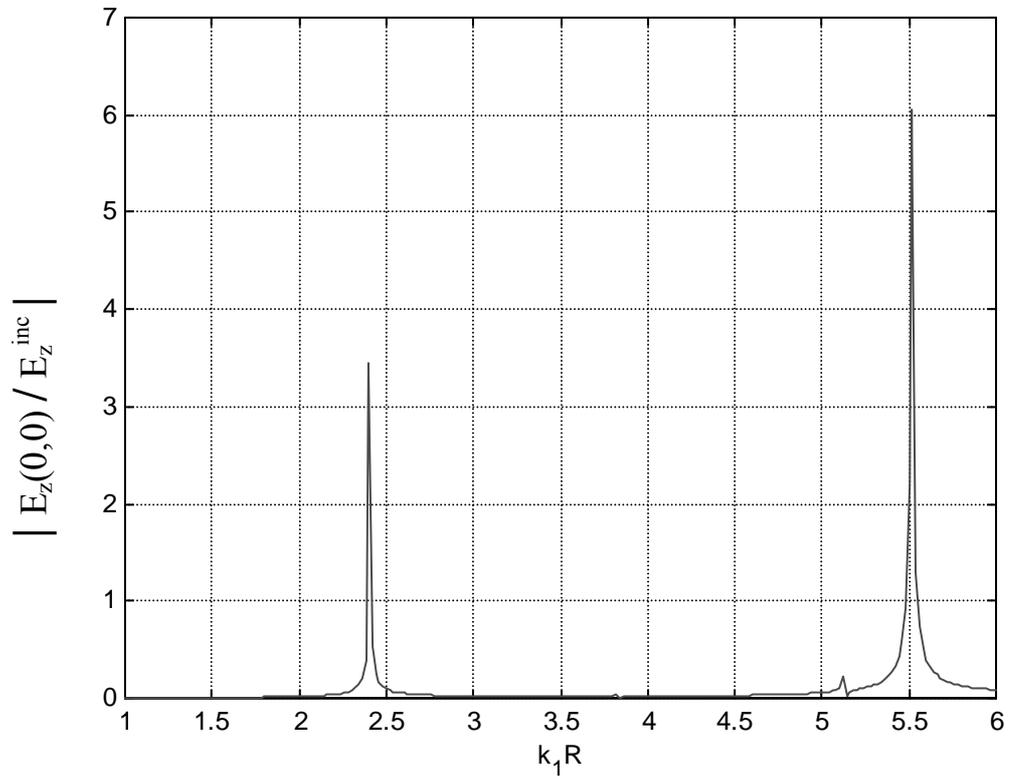
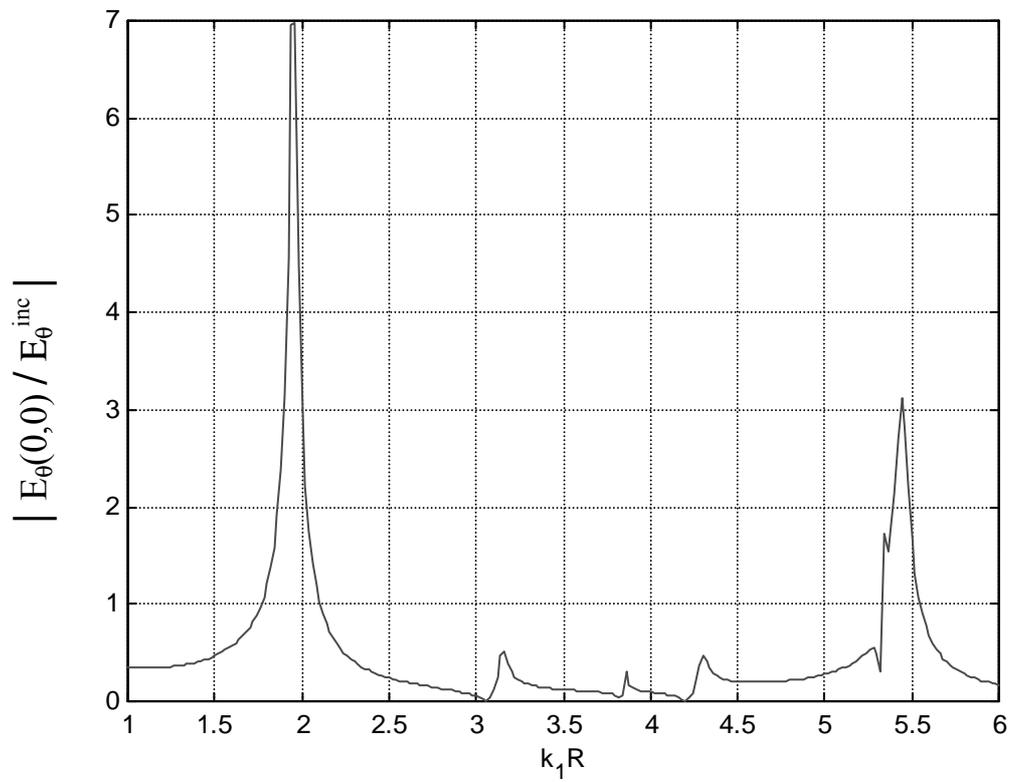
{\bf Figure 4} :(a) Normalized electric field amplitude at the center of the
cylinder for various

$\qquad \qquad \qquad k_{1}R$\ with $(\theta _{0}=0%
{{}^\circ}%
,\phi =5%
{{}^\circ}%
)$

\ \ \ \ \ \ \ \ \ \ \ \ \ \ \ (b) Normalized electric field amplitude at the
center of the cylinder for various

$\qquad \qquad \qquad k_{1}R$ with $(\theta _{0}=0%
{{}^\circ}%
,\phi =5%
{{}^\circ}%
)$

{\bf Figure 5} :\ Map of the electric field for values of $k_{1}R$
corresponding to the modes :

\qquad \qquad\ (a) TM$_{01}$and (b) TM$_{11}$of the cylinder. \newline
\newpage

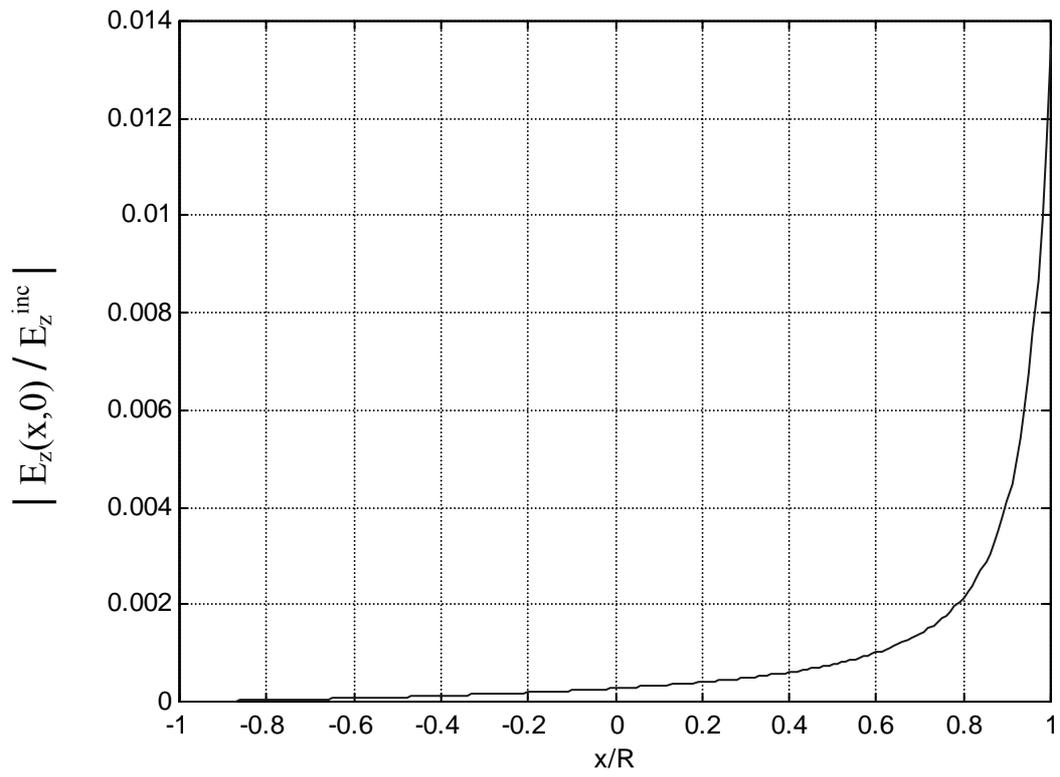
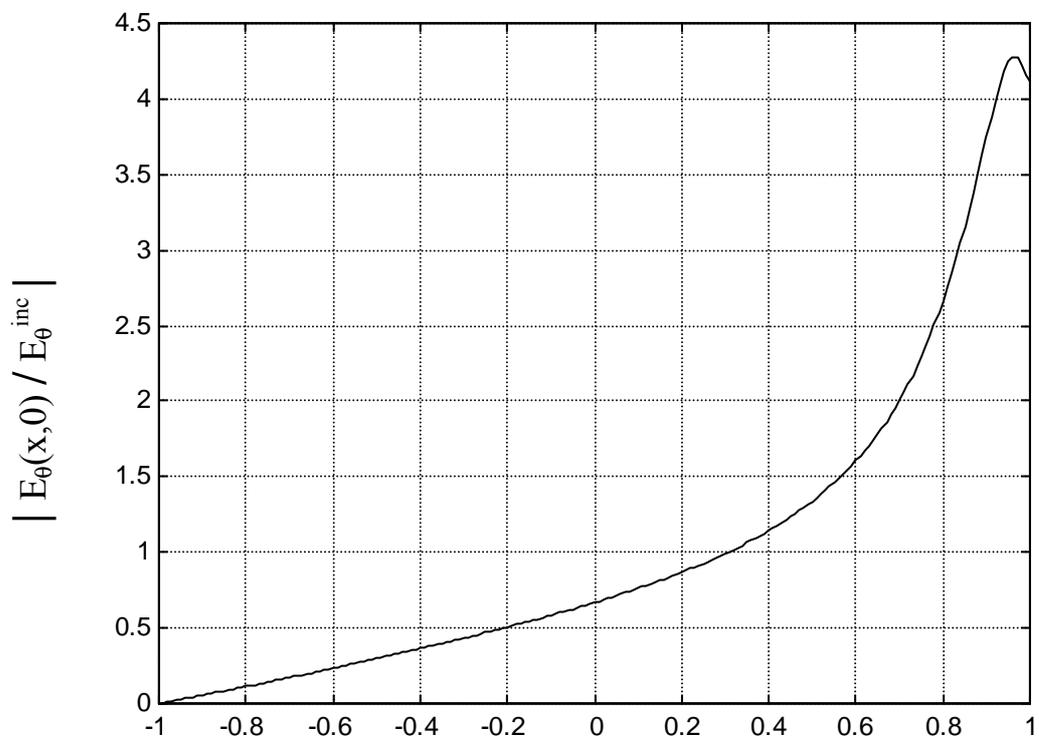
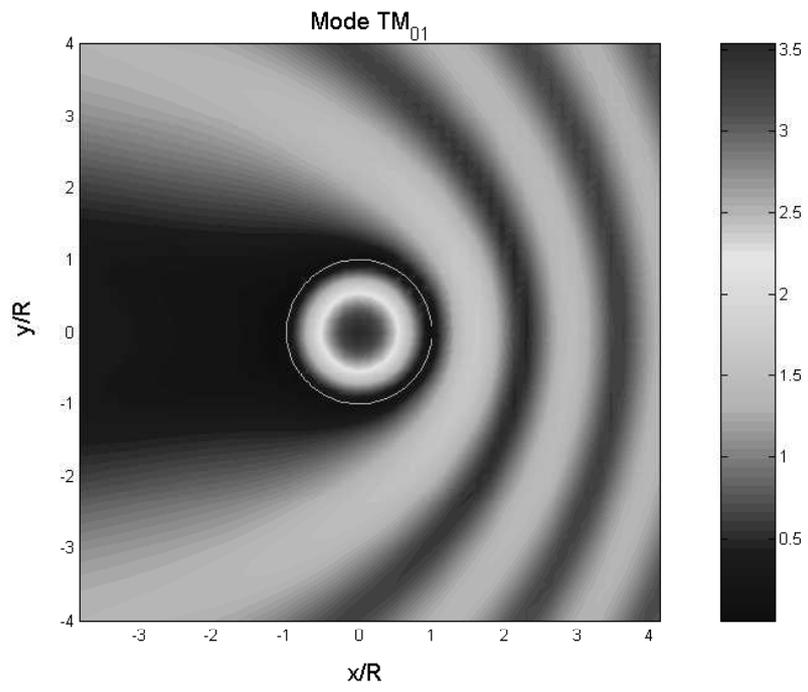
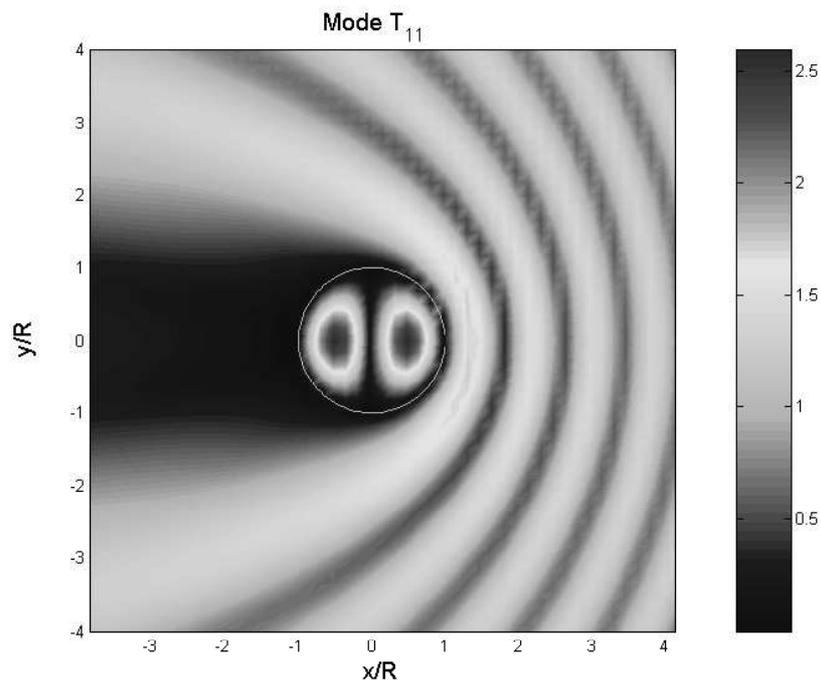
\end{document}